\documentstyle[sprocl,epsfig,wrapfig,rotating]{article}

\def\gg{\gamma\gamma}
\def\H{\rm h}
\def\WW{\mbox{WW}}
\def\ZZ{\mbox{ZZ}}
\def\g{\mbox{g}}
\def\h{\mbox{h}}
\def\ccbar{\overline{\mbox c}\mbox{c}}
\def\bbbar{\overline{\mbox b}\mbox{b}}
\def\qqbar{\overline{\mbox q}\mbox{q}}
\def\ccbarg{\overline{\mbox c}\mbox{cg}}
\def\bbbarg{\overline{\mbox b}\mbox{bg}}
\def\BR{\rm BR}

\newcommand{\AmS}{{\protect\the\textfont2
  A\kern-.1667em\lower.5ex\hbox{M}\kern-.125emS}}

\begin{document}


\markboth{FREIBURG-EHEP-99-08}{FREIBURG-EHEP-99-08}

\setcounter{footnote}{0}
\renewcommand{\thefootnote}{\fnsymbol{footnote}}
\title{LIGHT HIGGS PRODUCTION AT THE COMPTON COLLIDER\footnotemark[1] }

\author{GEORGI JIKIA and STEFAN S\"OLDNER-REMBOLD}

\address{
        Albert--Ludwigs--Universit\"at Freiburg,\\
        Hermann--Herder--Str.\ 3, 
        D--79104 Freiburg, Germany}

\maketitle\abstracts{
We have studied the production of a light Higgs boson with a mass of 120 GeV
in photon-photon collisions at a Compton collider.  The event
generator for the backgrounds to a Higgs signal due to $\bbbar$ and 
$\ccbar$ heavy quark pair production in polarized $\gamma\gamma$
collisions is based on a complete next-to-leading order (NLO) perturbative
QCD calculation. For $J_z=0$ the large double-logarithmic corrections
up to four loops are also included.  It is shown that the
two-photon width of the Higgs boson 
can be measured with high statistical accuracy of about 2\% for
integrated $\gg$ luminosity in the hard part of the spectrum of
$40$~fb$^{-1}$. As a result the total Higgs boson width can be calculated
in a model independent way to an accuracy of about 14\%.
}


\section{Introduction}
\setcounter{footnote}{0}
\renewcommand{\thefootnote}{\fnsymbol{footnote}}
\footnotetext[1]{
submitted to the proceedings of the International Workshop on Linear Colliders (LCWS99) at Sitges, Spain, 28 April - 5 May 1999 }
The experimental discovery of the Higgs boson is crucial for the
understanding of the mechanism of electroweak symmetry breaking. The
search for Higgs particles is one of the main goals for LEP2 and
Tevatron and will be one of the major motivations for the future Large
Hadron Collider (LHC) and Linear e$^+$e$^-$ Collider (LC). Once the
Higgs boson is discovered, it will be of primary importance to
determine its tree-level and one-loop
induced couplings, spin, parity, $CP$-nature, and its
total width in a model independent way. 
In this respect, the $\gamma\gamma$ Compton 
Collider~\cite{Telnov} option of LC offers a unique possibility to produce the
Higgs boson as an $s$-channel resonance decaying into $\bbbar$,
WW$^*$ or ZZ:
$$
\gamma\gamma\to \h^0\to \bbbar  \qquad (\WW^*, \ZZ)
$$
and thereby to measure the two-photon Higgs width. This partial width
is of special interest, since it first appears at the one-loop level so
that all heavy charged particles which obtain their masses from
electroweak symmetry breaking contribute to the loop. Moreover, the
contributions of very heavy particles do not decouple. In addition,
combined measurements of $\Gamma(\h\to\gamma\gamma)$ and
$\BR(\h\to\gamma\gamma)$ at the LC provide a model independent
measurement of the total Higgs width~\cite{snow96}.

The lower bound on $m_{\rm h}$ from
direct searches at LEP is $95.2$~GeV at 95 $C.L.$~\cite{tampere}. 
Recent global analysis of precision electroweak data~\cite{Mnich}
suggests that the Higgs boson is light $m_{\rm h}=92^{+78}_{-45}$~GeV. 
This
fact is in a remarkable agreement with the well known upper bound of
$\sim 130$ GeV for the lightest Higgs boson mass in the minimal
version of supersymmetric theories, the Minimal Supersymmetric
Standard Model (MSSM)~\cite{MSSM}. For this case of a light Higgs
boson the results of the Monte Carlo simulations of the Higgs
production in $\gamma\gamma$ collisions with final decay to $\bbbar$
quark pairs will be presented here. Taking into
account that the current upper bound on the Higgs mass from radiative
corrections is $m_{\rm h}<245$~GeV at 95\% $C.L.$ \cite{Mnich}, one can still 
hope to measure the two-photon Higgs width at the 300-500 GeV LC for heavier
Higgs masses by studying its production in $\gamma\gamma$
collisions and final decays into $WW^*$~\cite{WW*} or $ZZ$~\cite{ZZ}
states.

The accuracy of the $\Gamma(\h\to\gamma\gamma)$ measurements to be reached
can be inferred from the results of the studies of the coupling of the
lightest SUSY Higgs boson to two photons in the decoupling regime
\cite{decoupling}. It was shown that in the decoupling limit, where all
other Higgs bosons are very heavy and no supersymmetric particle has
been discovered at LHC or LC, chargino and top squark loops can
generate a sizable difference between the standard and the SUSY
two-photon Higgs couplings. Typical deviations are at the few percent
level. Top squarks heavier than 250 GeV can induce deviations even
larger than $\sim 10\%$ if their couplings to the Higgs boson are large.

\section{Signal and Background}

The cross-section of the resonant Higgs production at $\gamma\gamma$
Collider is proportional to the product
\begin{equation}
\sigma(\gamma\gamma\to \h^0\to X) = z \frac{dL_{\gamma\gamma}}{dz}
\frac{4\pi^2}{M_{h^0}^3} \Gamma(\h^0\to\gamma\gamma)
\cdot \BR(\h^0\to X)(1+\lambda_1\lambda_2).
\end{equation}
Here the effective photon-photon luminosity $L_{\gamma\gamma}$ is
introduced (see the next section). $\lambda_{1,2}$ are mean high
energy photon helicities.

Standard Model (SM) Higgs branching ratios and Higgs total width are
calculated with the help of the program HDECAY \cite{HDECAY}. The
program includes the full massive NLO corrections for $h\to \qqbar$
decays close to the thresholds as well as the massless ${\cal
O}(\alpha_s^3)$ corrections far above the thresholds. For the Higgs
signal only two-particle final states are generated, so that Parton
Shower (PS) algorithm of JETSET have been used to simulate three and higher
particle final states.

The main background to the $h^0$ production is the continuum
production of $\bbbar$ and $\ccbar$ pairs. In this respect, the
availability of high degree of photon beams circular polarization is
crucial, since for the equal photon helicities $(\pm\pm)$ that produce
spin-zero resonant states, the $\gamma\gamma\to \qqbar$ QED Born
cross-section is suppressed by the factor $m_q^2/s$
\cite{BBC}:
\begin{eqnarray}
&&\frac{d\sigma^{\rm Born}(J_z=0)}{dt} =\frac{12 \pi \alpha^2 Q_q^4}{s^2}
\frac{m_q^2 s^2(s-2 m_q^2)}{t_1^2 u_1^2},
\end{eqnarray}
and
\begin{equation}
\frac{d\sigma^{\rm Born}(J_z=\pm2)}{dt} = 
\frac{12 \pi \alpha^2 Q_q^4}{s^2}
\frac{(t_1 u_1 - m_q^2 s)(u_1^2+t_1^2+2 m_q^2 s)}
{t_1^2 u_1^2}.
\end{equation}
Here $m_q$ is the quark mass, $Q_q$ its charge, and $t_1 = t-m_q^2$,
$u_1 = u-m_q^2$.

Virtual one-loop QCD corrections for $J_z=0$ were found to be
especially large due to the double-logarithmic enhancement factor, so
that the corrections are comparable or even larger than the Born
contribution for the two-jet final topologies~\cite{JikiaTkabladze}. 
For small values of the cutoff $y_{\rm cut}$
separating two and three-jet events two-jet cross-section, calculated
to order $\alpha_s$, becomes even negative in the central region.
Recently leading double-logarithmic QCD corrections for $J_z=0$ were
resummed to all orders \cite{Sudakov}. The account of non-Sudakov form
factor to higher orders makes the cross-section well defined and
positive definite in all regions of the phase space.

The simulation program includes exact one-loop QCD corrections to
heavy quark production in $\gamma\gamma$ collisions
\cite{JikiaTkabladze} and non-Sudakov form factor in the
double-logarithmic approximation through four loops \cite{Sudakov}:
\begin{eqnarray}
&&\frac{\sigma^{\rm DL}_{\rm virt}}{\sigma_{\rm Born}} \sim 1 + 6 {\cal F}
+\frac{1}{6} \left(56 +2 \frac{C_A}{C_F} \right) {\cal F}^2
+\frac{1}{90} \left( 94 +90 \frac{C_A}{C_F} +2 \frac{C_A^2}{C_F^2} \right)
{\cal F}^3 \nonumber \\
&& +\frac{1}{2520} \left( 418 + 140 \frac{C_A}{C_F} + 238 \frac{C_A^2}{C_F^2}
+3 \frac{C_A^3}{C_F^3} \right) {\cal F}^4 
\end{eqnarray}
where ${\cal F} = - C_F \frac{\alpha_s}{4 \pi} \log^2 \frac{m_q^2}{s}$
is the one-loop hard form factor. Since it is a non-trivial task to
write down an event generator including both NLO corrections and
PS algorithm, we do not use any PS for
background $\bbbar$ and $\ccbar$ production. So the experimental value
of $y_{\rm cut}$ parameter should not be taken very small, otherwise the
account of resummed Sudakov corrections is needed. Two parton ($\bbbar$,
$\ccbar$, and $\bbbarg$, $\ccbarg$ with $y_{\rm cut}<0.01$) and three
parton ($\bbbarg$, $\ccbarg$ with $y_{\rm cut}>0.01$) final states
were generated separately and JETSET \cite{PYTHIA} string
fragmentation algorithm was used afterwards. The event generator
itself both for the Higgs signal and heavy quark background is
implemented using the programs BASES/SPRING \cite{BASES}.

\section{\boldmath $\gamma\gamma$ \unboldmath Luminosity}

The original polarized photon energy spectra \cite{luminosity} were
used. 100\% laser and 85\% electron beam polarizations were assumed
with $2\lambda_e^{1,2}\lambda_\gamma^{1,2}=-0.85$. The Parameter
$x=\frac{4E_e\omega_0}{m_e^2}$ was taken to be equal to 4.8.  Assuming
that the Higgs boson will already be discovered at LHC and/or LC and
its mass will be also measured, we tune the ee collision energy to be
equal to 152~GeV so that the Higgs mass of 120~GeV is just at the
peak of the photon-photon luminosity spectrum
$z\frac{dL_{\gamma\gamma}}{dz}$, $z=0.8$, where 
$$z=\frac{W_{\gamma\gamma}}{2E_e}, \quad z_{\rm max}=\frac{x}{x+1}=0.83$$
We assume an integrated $\gamma\gamma$ luminosity of 
$L_{\gamma\gamma}(0<z<z_{\max}) = 150$~fb$^{-1}.$ 
Realistic simulations of the $\gamma\gamma$ luminosity \cite{Telnov}
taking into account beamstrahlung, coherent pair creation and
interaction between charged particles show that idealized spectra
\cite{luminosity} will be strongly distorted in the low energy part of
the spectrum. However, in the hard part of the spectrum which is relevant
for our simulation, the spectra \cite{luminosity} represent a
very good approximation \cite{Telnov}. The
luminosity in the hard part of the spectrum is
$L_{\gamma\gamma}(0.65<z<z_{\rm max}) = 43$~fb$^{-1}$ which 
corresponds to a luminosity of the $ee$ collisions of
$L_{\rm ee}\approx 200$~fb$^{-1}$.

\section{Cross-Sections}

\begin{table}[hbt]
\setlength{\tabcolsep}{1.5pc}
\newlength{\digitwidth} \settowidth{\digitwidth}{\rm 0}
\catcode`?=\active \def?{\kern\digitwidth}
\begin{tabular}{|r|c|c|c|}\hline
$m_{h^0}=120$ GeV
&$\sigma$ [pb]& 
$N_{\rm ev}$ & $N_{\rm ev}$ \\

&$|\cos\theta|)<0.7$& 
& with b-tag \\\hline 
$\gamma\gamma\to h^0\to \bbbar$ &
0.105 & 15700& 11000
\\\hline 
$\gamma\gamma\to \bbbar (\g)$, $J_z=0$ &
0.0294 & & 
\\
$J_z=2$ &
0.0654 & & 
\\
total &
0.0947& 14200&  9940
\\\hline 
$\gamma\gamma\to \ccbar (\g)$, $J_z=0$ &
0.425 & & 
\\
$J_z=2$ &
1.03 & & 
\\
total &
1.45 &217000 & 7610
\\\hline 
$\gamma g\to \bbbar$ &
$7.45\ 10^{-4}$& 112& 78
\\\hline 
$\gamma g\to \ccbar$ &
$2.96\ 10^{-3}$&445 & 16
\\\hline 
\end{tabular}
\caption{Cross-sections and event rates for the Higgs signal and
backgrounds from direct and resolved heavy quark production in
$\gamma\gamma$ collisions. The $\gamma\to\bbbar (\g),\ccbar (\g)$
background was simulated for $\gamma\gamma$ invariant masses
larger than 100 GeV.}
\end{table}

In Table 1, we give the cross-sections and event rates for the
Higgs signal and backgrounds calculated without detector simulation.
For $\bbbar (\g)$, $\ccbar (\g)$ backgrounds the $J_z=0$ and $J_z=2$
contributions are shown separately. b-tagging efficiencies of
70 \% for $\bbbar$ events and 3.5 \% for $\ccbar$ events were assumed.
Both quark jets were assumed to satisfy the $|\cos\theta|<0.7$ cut. 
The resolved photon
contribution to the $\bbbar$ background is negligible. It was therefore not
included in the subsequent detector simulation analysis. Assuming
$$
\frac{\Delta\Gamma(\h\to\gg)}{\Gamma(\h\to\gg)}= 
\frac{\sqrt{N_{\rm obs}}}{N_{\rm obs}-N_{\rm BG}},
$$
where $N_{\rm obs}$ is the sum of the signal and background events
and $N_{\rm BG}$ the number of background events,
the first estimate of the statistical error of two-photon Higgs
partial width measurement is 1.5\% for $m_h=120$~GeV. For heavier
Higgs masses we get the statistical error of 1.9\% (6.1\%) for
$m_h=140$~GeV ($m_h=160$~GeV), respectively.

\section{Results of the Monte Carlo Simulation}

\begin{figure}[htb]
\begin{center}
\epsfig{file=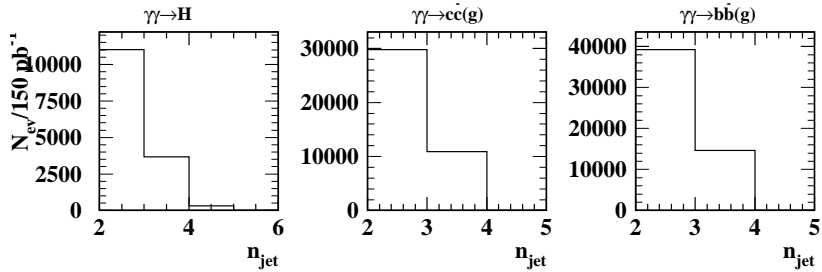,width=0.85\textwidth}
\label{fig:multiplicities}
\caption{Jet multiplicities for two- and three-parton events.}
\end{center}
\end{figure}

The Monte Carlo simulation of the fragmentation
is done with JETSET~\cite{PYTHIA}. Signal and
background are studied using the fast detector simulation program
SIMDET~\cite{SIMDET} for a typical TESLA detector. The Higgs mass was
assumed to be 120~GeV. Jets were reconstructed using the Durham algorithm
with $y_{\rm cut}=0.02$.  b tagging is not simulated and b tagging
efficiencies of 70 \% for $\bbbar$ events and 3.5 \% for $\ccbar$
events were used instead~\cite{Battaglia}. These efficiencies are based
on double-tagging of b jets to suppress $\ccbar$ events 
by a factor of 20 which is large enough to overcome the
enhancement factor of 16 due to the larger c quark charge. 
The
multiplicities for the Higgs signal and heavy quark background are
shown in Fig.~\ref{fig:multiplicities}.
The following cuts were used to suppress heavy quark background:

\noindent
1.) To suppress the background at the peak, exactly two jets
must be found in the event ($ n_{\rm jet}=2$).

\begin{figure}[htb]
\begin{center}
\epsfig{file=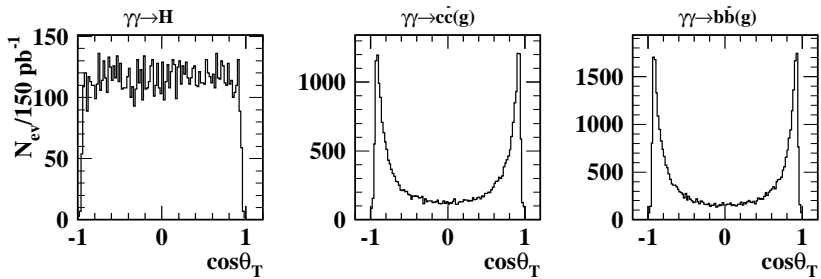,width=0.85\textwidth}
\end{center}
\caption{$\cos\theta_T$ distributions for two- and three-parton
events with $n_{\rm jet}=2$.}  
\label{fig:cos}
\end{figure}

\noindent 2.) Events where quark jets were scattered at a small angle were
rejected by requiring for the thrust angle $|\cos\theta_{\rm T}|<0.7$
(see Fig.~\ref{fig:cos}).



\begin{wrapfigure}[14]{r}{2.5in}
\begin{center}
\epsfig{file=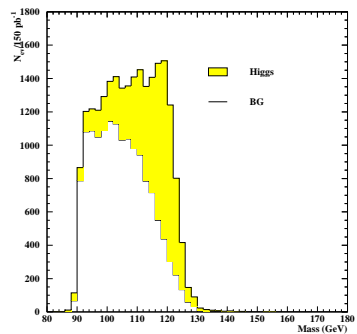,width=0.36\textwidth}
\end{center}
\label{fig:higgs}
\caption{Mass distributions for Higgs signal and heavy quark
background.}  
\end{wrapfigure}

\noindent 3.) Since the Higgs boson is produced at the peak of photon-photon
luminosity almost at rest the cut on longitudinal momentum component
of the event divided by the ee centre-of-mass energy,
$|p_{\rm z}|/\sqrt{s}_{\rm ee}<0.1$, further reduces the background.

\noindent 4.) Most background events are produced at the lower energy tail
of the photon-photon luminosity distribution. A cut on the total visible
energy, $E_{\rm vis}/\sqrt{s}_{\rm ee}>0.6$, eliminates most
soft background events.

After these cuts a signal efficiency of about 50\% is achieved. 
The invariant mass distributions for the combined $\bbbar (\g)$
and $\ccbar (\g)$ background, and for the Higgs signal 
are shown in Fig.~\ref{fig:higgs}.


After these cuts 8104 signal events and 14690 background events
remain yielding a relative error 
$$\frac{\Delta[\Gamma(\h\to\gg)\BR(\h\to
\bbbar)]}{[\Gamma(\h\to\gg)\BR(\h\to\bbbar)]}\approx 2\%.$$ 
Since resummation of Sudakov logarithms was not done
for the heavy quark background a question may arise on the effect of
these corrections on the errors on $\Gamma(\h\to\gg)$. 
To estimate this
effect the analysis was repeated without cut on $n_{\rm jet}$. 
In this case no Sudakov logarithms are present. 
This increases the signal by about 36\% and the background by about
50\% in Fig.~\ref{fig:higgs} which yields a comparable
statistical error as with jet multiplicity cut.
In comparison to the results presented in Table~1
taking into account of the detector simulation effects gives slightly lower
accuracy. Taking this value and assuming, that at the $e^+e^-$ linear
collider one can measure the $h\to \bbbar$ and $h\to\gg$ branching
ratios with the accuracy of
$$\frac{\Delta\BR(\H\to\bbbar) }{\BR(\H\to\bbbar) }=5\%\mbox{~~~and~~~}
\frac{\Delta\BR(\H\to\gg) }{\BR(\H\to\gg) }=13\%$$
one can estimate, that the total width of the Higgs boson can be calculated
$$\Gamma_{\H}=\frac{[\Gamma(\H\to\gg) \BR (\H\to\bbbar)]}{
[\BR(\H\to\gg)] [\BR(\H\to\bbbar)]}$$ 
to the accuracy of about 14\%. This error is dominated by the expected
error on $\BR(\h\to\gamma\gamma)$.

The influence of the values of the b tag efficiencies for 
$\bbbar$ and $\ccbar$ events on
the accuracy of two-photon Higgs width was also studied.
b and c tag efficiencies were taken from 
a parametrisation by M.~Battaglia \cite{Battaglia}.
In the region of b tag efficiencies from 50\% to 90\% the relative
error on $\Gamma(\h\to\gg)\BR(\h\to \bbbar)$ is quite stable and has a
value about 2\%.

\section{Conclusions}

Our preliminary results show that the two-photon width of the Higgs
boson can be measured at the photon-photon collider with high statistical 
accuracy
of about 2\%. At such an accuracy one can discriminate between the SM
Higgs particle and the lightest scalar Higgs boson of the MSSM in the
decoupling limit, where all other Higgs bosons are very heavy and no
supersymmetric particle has been discovered at the $e^+e^-$ LC. Due to
the large charm production cross-section in $\gamma\gamma$ collisions,
excellent b tagging is required.  Our results are consistent with
the earlier results \cite{Watanabe}, where higher order double
logarithmic correction were not taken into account. To get more
realistic $\Gamma(\h\to\gg)$ error one still needs to understand the
systematic errors. Since precise knowledge of the heavy quark
background is essential for accurate measurements of the Higgs signal,
it is worth mentioning, that one should not necessarily rely on the
theoretical predictions of $\bbbar$ and $\ccbar$ backgrounds. Since
the light Higgs boson is expected to be very narrow,
it will be possible by slightly
changing the electron beam energy to completely eliminate the Higgs
signal events and thereby just to measure the heavy quark background.

\section*{Acknowledgments}
The valuable discussions with I.F.~Ginzburg, R.~Hawkings, 
V.A.~Khoze, M. Melles,
V.I.~Telnov are gratefully acknowledged.


\section*{References}
\begin{thebibliography}{9}
\bibitem{Telnov} V.I.~Telnov, these proceedings.
\bibitem{Mnich} J.~Mnich, plenary talk at the Int.
Europhysics Conf. on High Energy Physics, Tampere, Finland,
15--21 July, 1999.
\bibitem{tampere}
The LEP collaborations, paper submitted to 
the Int.
Europhysics Conf. on High Energy Physics, Tampere, Finland,
15--21 July, 1999.
\bibitem{MSSM}S. Heinemeyer, W. Hollik, G. Weiglein, Eur. Phys. J. C9
(1999) 343; 
H E. Haber, to be published in the proceedings of 4th Int.
Symposium on Radiative Corrections, Barcelona, Spain, 8-12
September 1998, hep-ph/9901365.
\bibitem{WW*} E.~Boos, V.~Ilyin, D.~Kovalenko, T.~Ohl, A.~Pukhov, M.~Sachwitz, 
H.J.~Schreiber, Phys. Lett. B427 (1998) 189.
\bibitem{ZZ} G.~Jikia, Phys.Lett. B298 (1993) 224;
Nucl. Phys.  B405 (1993) 24;
M.S.~Berger Phys. Rev.  D48 (1993) 5121;
D.A.~Dicus and C.~Kao, Phys. Rev.  D49 (1994) 1265.
\bibitem{snow96}
J.F.~Gunion, L.~Poggioli, R.~Van~Kooten, C.~Kao, P.~Rowson, {\it
Proc. of the 1996 DPF/DPB Summer Study on ``New Directions in
High Energy Physics'' (Snowmass, 96)}, June 25 - July 12, 1996,
Snowmass, Colorado, March, 1997, UCD-97-5, hep-ph/9703330.
\bibitem{decoupling} A. Djouadi, V. Driesen, W. Hollik, J.I. Illana,
Eur. Phys. J. C1 (1998) 149.
\bibitem{luminosity} I.F.~Ginzburg, G.L.~Kotkin, S.L.~Panfil, V.G.~Serbo, 
V.I.~Telnov, Nucl. Instr. Meth. 219 (1984) 10.
\bibitem{HDECAY} A.Djouadi, J.Kalinowski and M.Spira,
Comp. Phys. Comm. 108 (1998) 56.
\bibitem{BBC} D. Borden, D. Bauer, D. Caldwell, {\it SLAC-PUB}-5715,
{\it UCSD-HEP}-92-01.
\bibitem{JikiaTkabladze} G. Jikia, A. Tkabladze,
Proc. of the {\it Workshop on  gamma--gamma colliders}, March 28-31,
1994, Lawrence Berkeley Laboratory, Nucl. Instr. Meth.
A355 (1995) 81;
Phys. Rev.  D54 (1996) 2030;
\bibitem{Sudakov} M. Melles, W.J. Stirling, Phys. Rev. D59 (1999) 94009; 
Eur. Phys. J. C9 (1999) 101; 
M. Melles, W.J. Stirling, V.A. Khoze, hep-ph/9907238;
M. Melles, these proceedings.
\bibitem{BASES} S.~Kawabata, Comp. Phys. Comm. 88 (1995) 309.
\bibitem{PYTHIA} T.~Sj\"ostrand, Comp. Phys. Comm. 82 (1994) 74. 
\bibitem{SIMDET} M.~Pohl, H.J.~Schreiber, 
DESY-99-030. 
\bibitem{Battaglia} M.~Battaglia, private communications.
\bibitem{Watanabe}I.~Watanabe et. al., KEK-REPORT-97-17, March 1998.
\end{thebibliography}
\end{document}